\documentclass[12pt]{iopart}

\begin{document}

\title[Non-Gaussian distributions under scrutiny]{Non-Gaussian distributions under scrutiny}

\author{Thierry Dauxois}

\address{Laboratoire de Physique, ENS Lyon, CNRS, 46 All\'{e}e
d'Italie, 69364 Lyon cedex 07, France}
\ead{Thierry.Dauxois@ens-lyon.fr}


\bigskip

The importance of the Gaussian distribution as a quantitative model of
stochastic phenomena is familiar to physicists. Brownian motion is
presumably the paradigmatic example; in this case, it is well known
that the sum of a very large number of small displacements is Gaussian
distributed with a variance that grows with time. From
the mathematical point of view, this result is made precise by the
central limit theorem, and is essentially valid provided that the
elementary displacements are sufficiently {\em decorrelated} one from
another. This explains, incidentally, why one usually cites the
drunkard's walk as an example of Brownian motion: it is a walk for
which decorrelation is provided by the wine!

The Gaussian is not the only limit distribution for sums
of random variables. Numerous examples of non-Gaussian distributions
have been reported in the context of what is nowadays loosely called
``complex systems'', and which include disordered systems, systems
undergoing phase transitions, turbulent fluids, astrophysical systems,
finance time series, social networks, etc. In these systems, sums of
random variables can be defined, but the correlations between the
variables are then so strong that they cannot be omitted, with the end
result that the distribution of the sum is not, in general, Gaussian
distributed in the limit of infinite number of events. Many different
approaches to predict what the limiting distribution is have been
proposed in complicated problems involving long-range
interactions, memory effects, etc.

In this context, it has been suggested that the so-called
$q$-Gaussian distribution, defined by
\begin{equation}G_q(x) = A(1-(1 - q)\beta x^2)^{1/1-q},\label{qgaussian}\end{equation}
could be the basis for a generalized central limit theorem. In this
expression, $A$ is a normalization constant and $\beta$ controls the
width of the distribution. The basis for this suggestion~\cite{tsallis} is that the
distribution (\ref{qgaussian}) maximizes
\begin{equation}
S_q=\frac{\displaystyle 1-\int p(x)^q\ \mbox{d}x}{q-1}
,\label{entrotsallis}
\end{equation}
which would reduce to the usual Shannon entropy $-\int p(x)\ln p(x)
\mbox{d}x$ when the so-called entropic index tends to 1. 
Expression~(\ref{entrotsallis}) would thus be a generalization of
statistical mechanics, often called non extensive statistical
mechanics.  It is important to emphasize that the parameter $q$ allows
for an interpolation between the ``window'' function which is constant
over a finite support (obtained for $q\rightarrow -\infty$), the
ordinary Gaussian distribution ($q = 1$), and distributions having
power-law tails ($1 < q < 3$). The function $G_q(x)$ has therefore
quite a large fitting spectrum and might be applied for the study of
strongly correlated systems.

Independently of their applicability to statistical mechanics,
$q$-Gaussians have impressive mathematical properties and became
rapidly popular. They have been proposed to describe numerous
experimental or numerical results: velocity distributions of classical
rotators or galaxy clusters, turbulent flows, cellular aggregates or
the temperature fluctuations in the cosmic microwave background,...
Unfortunately, in the absence of firm grounds, physicists have
distributed themselves between enthusiasts and skeptics. At this point
in time, it is therefore important to distinguish whether the
formalism suggested by~(\ref{qgaussian}) and~(\ref{entrotsallis}) can
be the basis for developing a real predictive theory or if it is
``just'' a nice idea and a powerful fitting function. Two important
questions are particularly pressing here: (i) does the $q$-Gaussian
law describe the details of some physical problems and, more
importantly, (ii) is anyone able to provide analytical predictions of
the value of the $q$-index in terms of the microscopic parameters of
the physical system.

A particularly interesting paper in this respect is the one by Henk
Hilhorst and Gregory Schehr~\cite{Hilhorst}. It is an important
step for statistical physics since the authors are able to show by
explicit calculations that, in two examples of random variables,
previously put forward as candidates for being $q$-Gaussian
distributed, the probability distributions of the sums turn out to
be analytically different, although they closely resemble them
numerically.

The first example is presumably the simplest imaginable instance of
a strongly correlated system~\cite{Thistleton}, namely the scaled
sum $\sum_j u_j/N$, where the $N$ variables $u_j$ are identically
distributed on a finite domain, but with strong mean-field
correlations. For this example, Thistleton {\em et al} conjectured
from numerical fits that the distributions of the sums is
$q$-Gaussian. Hilhorst and Schehr, however, show
analytically~\cite{Hilhorst} that it is not. 

The second model~\cite{Moyano} consists of $N$ Boolean random
variables, correlated in an implicit way. For this model
$q$-Gaussians were also observed numerically, but Hilhorst and
Schehr~\cite{Hilhorst} show again that the true distribution
underlying the model is not a $q$-Gaussian. Yet, as it is nicely
stated in Hilhorst and Schehr paper, ``things conspire again such
that it becomes extremely difficult to distinguish the true curve
from its $q$-Gaussian approximant''. In the end, although the
definition of the model was clearly motivated by the formalism of
$q$-Gaussians, Hilhorst and Schehr show that $q$-Gaussians do not
pass a careful inspection.

These considerations are rather reminiscent of recent works on
experimental Lagrangian turbulence. After a preliminary and promising
study~\cite{Beck}, which showed that the distribution of accelerations
in turbulent fluids could be fitted by $q$-Gaussians, an improved set
of experimental data~\cite{Mordant} later showed that they were, after
all, neither well described nor well fitted by $q$-Gaussians.

In summary, notwithstanding the interesting properties exhibited by
$q$-Gaussians, the two examples studied in Ref.~\cite{Hilhorst} do not
lend support to the idea that these functions play any special role as
limit distributions of correlated sums. Clearly, more work is called
for establishing a more comprehensive picture and to satisfactorily
assess the role (if any) of $q$-Gaussians in statistical
mechanics. 
The future is open, but if there is one lesson that has to
be learned here, it is that one should be extremely careful when
interpreting non-Gaussian data in terms of $q$-Gaussians.

\end{document}